\documentclass[aps,prl,twocolumn,floats,graphicx,showpacs]{revtex4}
\usepackage{bbm}
\usepackage{amssymb}
\usepackage{ifpdf}
\usepackage{graphicx,epsfig}
\newcommand{\Z}{{\mathbb{Z}}}
\newcommand{\R}{{\mathbb{R}}}

\newcommand{\p}{\partial}
\begin{document}
\title{The Width of the Confining String in Yang-Mills Theory}
\author{F.\ Gliozzi$^a$, M.\ Pepe$^b$, and U.-J.\ Wiese$^c$}
\affiliation{
$^a$ Dipartimento di Fisica Teorica, Universit\`a di Torino, and INFN, Sezione 
di Torino, via P.\ Giuria 1, 10125 Torino, Italy \\
$^b$ INFN, Sezione di Milano-Bicocca, Edificio U2, Piazza della Scienza 3,
20126 Milano, Italy \\
$^c$ Albert Einstein Center for Fundamental Physics,
Institute for Theoretical Physics, Bern University,
Sidlerstrasse\ 5, 3012 Bern, Switzerland}


\begin{abstract}

We investigate the transverse fluctuations of the confining string connecting 
two static quarks in $(2+1)$-d $SU(2)$ Yang-Mills theory using Monte Carlo 
calculations. The exponentially suppressed signal is extracted from the large 
noise by a very efficient multi-level algorithm. The resulting width of the 
string increases logarithmically with the distance between the static quark 
charges. Corrections at intermediate distances due to universal higher order 
terms in the effective string action are calculated analytically. They 
accurately fit the numerical data.
\end{abstract} 

\pacs{11.15.Ha, 12.38.Aw, 12.38.Gc, 12.38.Lg}

\maketitle

Understanding the dynamics of confining strings connecting static quarks and 
anti-quarks in non-Abelian gauge theories is a great challenge in strong 
interaction physics. During its time-evolution, a confining string sweeps out a 
world-sheet whose boundary is determined by the world-lines of the static 
external quark charges. The string world-sheet can also be viewed as an analog 
of a fluctuating interface separating different phases of condensed matter. 
Just like a rough interface, the confining string 
in a non-Abelian gauge theory supports massless transverse fluctuations.
Interestingly, the dynamics of these fluctuations --- known as capillary waves 
in condensed matter physics --- is captured by a systematic 2-dimensional 
low-energy effective field theory. In the effective theory, the string is 
described by a $(d-2)$-component vector $\vec h(x,t)$ pointing to the location 
of the string world-sheet in the $(d-2)$ transverse dimensions of a 
$d$-dimensional space-time. Here $x \in [0,r]$ and $t \in [0,\beta]$ are the 
Euclidean coordinates parameterizing the base-space, with $r$ being the fixed 
distance between the external static charges and the inverse temperature 
$\beta$ being the extent of Euclidean time. Since the string ends at the static 
quark charges, its fluctuation field obeys the boundary condition 
$\vec h(0,t) = \vec h(r,t) = 0$. The leading term in the action of the 
effective theory
\begin{equation}
\label{action}
S[\vec h] = \frac{\sigma}{2}  \int_0^\beta dt \int_0^r dx \ 
\p_\mu \vec h \cdot \p_\mu \vec h, \quad \mu \in \{1,2\},
\end{equation}
gives rise to a universal contribution to the static quark potential
\begin{equation}
\label{potential}
V(r) = \sigma r - \frac{\pi (d-2)}{24\, r} + {\cal O}(1/r^3),
\end{equation}
where $\sigma$ is the string tension. The universal L\"uscher term is 
proportional to the number $(d-2)$ of transverse directions in which the string 
is fluctuating \cite{Lue80,Lue81}. The effective action from above also 
accounts for the string width \cite{Lue81a}. Due to its transverse 
fluctuations, the string broadens as the quark sources are separated. More 
precisely the transverse area swept out by the flux tube increases 
logarithmically with separation. At the distance $r/2$ half-way between the 
external static quark sources, at leading order the string has the squared width
\begin{equation}
\label{width}
w^2_{lo}(r/2) = \frac{d-2}{2 \pi \sigma} \log \left(\frac{r}{r_0}\right),
\end{equation}
where $r_0$ is some distance scale. This fundamental formula for the 
width is again universal. It applies to strings in confining gauge theories, to 
fundamental strings, as well as to fluctuating interfaces in condensed matter 
physics. Interestingly, until now the logarithmic behavior of the string width 
has been verified only in the context of spin models. Below the critical 
temperature of the 3-d Ising model, but above the corresponding roughening 
transition, an interface separating the broken phases indeed has massless 
fluctuations described by the effective theory of eq.(\ref{action}). Using
an efficient cluster algorithm, the expression for the width of 
eq.(\ref{width}) has been verified in \cite{has}. By a duality transformation, 
the results obtained in the 3-d Ising model also apply to $(2+1)$-d Abelian 
$\Z(2)$ lattice gauge theory, where the effective theory has been confirmed in 
more detail also using different boundary conditions \cite{cas96}. Similar 
results have been obtained in a $(2+1)$-d $\Z(4)$ gauge theory \cite{giu07}. In
non-Abelian gauge theories the measurement of the string width is 
computationally very challenging; early calculations were affected by large 
statistical errors and did thus not lead to definite conclusions~\cite{bali95}.

In this paper, for the first time we verify the logarithmically divergent 
string width for a non-Abelian gauge theory by simulating $(2+1)$-d $SU(2)$ 
lattice Yang-Mills theory using a very efficient multi-level algorithm 
\cite{Lue01,Lue02}. As we will see, the Monte Carlo data agree with the 
analytic prediction at large distances $r$. Before we turn
to the numerical results, we systematically work out corrections at intermediate
distances, which arise due to higher order terms in the low-energy effective 
action. For $d = 3$ there is only one transverse dimension, and hence in this 
case $h(x,t) \in \R$ is a one-component scalar field. The leading and 
next-to-leading terms in the low-energy effective action describing the massless
transverse fluctuations of the string are then given by
\begin{equation}
\label{action3d}
S[h] = \frac{\sigma}{2} \int_0^\beta dt \int_0^r dx \  \left[\p_\mu h\, \p_\mu h -
\frac{1}{4} \left(\p_\mu h \, \p_\mu h\right)^2 \right].
\label{ngaction}
\end{equation}
Since the Dirichlet boundary conditions $h(0,t) = h(r,t) = 0$ explicitly break 
translation invariance, one may have expected boundary terms to be present in 
the effective action as well \cite{Lue02}. Remarkably, due to open-closed 
string duality, such terms are absent and the pre- factor of the first 
sub-leading term is uniquely determined \cite{Lue04}. A generalization of these
arguments shows that for $d = 3$ there is only one six-derivative term which
coincides with the one of the Nambu-Goto action \cite{aha09}.

The effective action of eq.(\ref{action3d}) describes string fluctuations in 
the continuum. Before one reaches the continuum limit, the confining string in 
a lattice Yang-Mills theory is also affected by lattice artifacts. First of all,
at very strong coupling the world-sheet swept out by the lattice string is
rigid, i.e.\ it follows the discrete lattice steps and does not even have
massless excitations. Only at weaker coupling, after crossing the roughening
transition, the string world-sheet supports massless excitations and thus 
becomes rough. Consequently, the effective theory is applicable only in the 
rough phase. 

For a string world-sheet with periodic boundary conditions in Euclidean 
time, the squared width of the string is given by
\begin{equation}
w^2(x)= \langle h(x,t)^2 \rangle,
\end{equation}
which is directly related to the two-point function 
$\langle h(x,t)\,h(x',t')\rangle$, with the two points $(x,t)$ and 
$(x',t')$ falling on top of each other. This limit leads to ultraviolet 
divergences which we regularize using the point-splitting procedure
\begin{equation}
\langle h(x,t)^2 \rangle \longrightarrow 
\langle h(x,t)\,h(x'=x+\epsilon,t'=t+\epsilon')\rangle.
\label{ps}
\end{equation}
At the end, one sets $\epsilon\to0\,,\epsilon'\to 0$ and the remaining ultraviolet
divergent terms are absorbed in physical length scales. In the leading order Gaussian
approximation of eq.(\ref{action}), the two-point function can be conveniently expressed
in the form
\begin{equation}
\langle h(x,t)\,h(x',t')\rangle=
\frac{1}{\pi \sigma}
\sum_{n=1}^\infty\sin n\xi_1
\sin n\xi_1' 
\frac{\mbox{e}^{-n\xi_2  } +q^n \mbox{e}^{n\xi_2}  }{n(1-q^n)},
\label{cylinder}
\end{equation}
where $\xi_1=\pi x/ r$, $\xi_1'=\pi x'/ r$, $\xi_2=\pi (t-t')/r$ with $0 \le \xi_2 \le \pi\beta/r$. 
The parameter $q=\mbox{e}^{2 \pi i \tau}$ depends on the ratio 
$\tau = i \beta/(2 r)$. One then obtains the squared width of the string as
\begin{equation}
w^2_{lo}(r/2)=\frac1{2\pi\sigma}\log\left(\frac r{r_0}\right) 
+\frac1{\pi\sigma}\log\left( \frac{\eta(2\tau)}{\eta(\tau)^2}\right),
\label{pss}
 \end{equation}
with $r_0 = \pi\vert\epsilon+i\epsilon'\vert/2$ and the Dedekind function
\begin{equation}
\eta(\tau)=q^{\frac1{24}}
\prod_{n=1}^\infty(1-q^n).
\end{equation}
In the numerical simulations we put $r \ll \beta$. Then there are only
exponentially small corrections to the logarithmic broadening of the width of 
the string. On the other hand,  for $r \gg \beta$, the inversion transformation 
property 
\begin{equation}
\eta(\tau)=\eta\left(-1/\tau\right)/\sqrt{-i\tau}
\end{equation}
implies that
\begin{equation}
w^2_{lo}(r/2)=\frac1{2\pi\sigma}\log\left(\frac{\beta}{4r_0}\right)+
\frac{r}{4\beta\sigma} + {\cal O}\left(\mbox{e}^{-2\pi r/\beta}\right).
\end{equation}
This shows that, at finite temperature, the confining string broadens linearly 
with the distance between the external static quarks \cite{Cas09}.

The non-Gaussian correction due to the first sub-leading term in 
eq.(\ref{ngaction}) results from the six-point function
\begin{equation}
\frac{\sigma}{8}\,
\left\langle
 h(\frac r2,0)\,h(\frac r2+\epsilon,\epsilon')\left[\int_0^rdx\int_0^{\beta}dt \;
(\p_\mu h \, \p_\mu h)^2\right] \right\rangle~.
\end{equation} 
This amplitude has two kinds of ultraviolet divergent terms: one proportional 
to $\log\vert\epsilon+i\epsilon'\vert$ and the other proportional to
$(\epsilon^2-\epsilon'^2)/\vert\epsilon+i\epsilon'\vert^4$. The former is 
absorbed in the scale of the logarithm and the latter is eliminated by putting
$\epsilon^2=\epsilon'^2$. Therefore the contribution to $w^2$ of the 
next-to-leading term does not require the introduction of new low-energy 
parameters. Up to first order in the expansion parameter $1/(\sigma r^2)$, the 
complete expression turns out to be
\begin{eqnarray}
\label{widthcorr}
&&\!\!\!\!\!\!\!\!\!w^2(r/2)=\left( 1 +\frac{4\pi f(\tau)}{\sigma r^2}\right) w^2_{lo}
(r/2)-\frac{f(\tau)+g(\tau)}{\sigma^2r^2}, \nonumber \\
&&\!\!\!\!\!\!\!\!\!f(\tau)=\frac{E_2(\tau)-4E_2(2\tau)}{48}, \nonumber \\
&&\!\!\!\!\!\!\!\!\!g(\tau)=i \pi \tau \left(\frac{E_2(\tau)}{12}- q\frac {d}{dq} \right) 
\left( f(\tau)+\frac{E_2(\tau)}{16} \right)+\frac{E_2(\tau)}{96}, \nonumber \\
&&\!\!\!\!\!\!\!\!\!E_2(\tau)=1-24\sum_{n=1}^\infty\frac{n\,q^n}{1-q^n}.
\end{eqnarray}
Here $E_2(\tau)$ is the first Eisenstein series. Just like the action, 
the operator $h(x,t)$ also receives higher-order corrections. Remarkably, their
effect vanishes in eq.(\ref{widthcorr}). The derivation of these
results will be presented in~\cite{gpw}.

Lattice gauge theory is a powerful tool that allows us to address the question
of string dynamics from first principles. Since it is least problematical for
numerical simulations, we consider $(2+1)$-d $SU(2)$ Yang-Mills theory on a 
cubic lattice of size $L \times L \times \beta$, with the Euclidean time extent
$\beta$ determining the inverse temperature. We use the standard Wilson 
plaquette action 
\begin{equation}
S[U] = - \frac{2}{g^2} \sum_{x,\mu,\nu} 
\mbox{Tr} [U_{x,\mu} U_{x+\hat\mu,\nu} U_{x+\hat\nu,\mu}^\dagger U_{x,\nu}^\dagger],
\end{equation}
for parallel transporter variables $U_{x,\mu} \in SU(2)$ in the fundamental 
representation of the gauge group, located on the links $(x,\mu)$. Here $g$ is 
the bare gauge coupling and $\hat\mu$ is a unit-vector pointing in the 
$\mu$-direction. It should be noted that all physical quantities are measured in
units of the lattice spacing which we put to 1. The partition function takes 
the form
\begin{equation}
Z = \int {\cal D}U \ \mbox{e}^{- S[U]} = \prod_{x,\mu} \int_{SU(2)} dU_{x,\mu} \
\mbox{e}^{- S[U]},
\end{equation}
where $dU_{x,\mu}$ denotes the local gauge invariant Haar measure on the link
$(x,\mu)$. An external static quark located at the site $x$ is represented by a Polyakov 
loop
\begin{equation}
\Phi_x =\frac{1}{2} \mbox{Tr} \left[\prod_{t=1}^\beta U_{x+t\hat 2,2}\right],
\end{equation}
a parallel transporter wrapping around the periodic Euclidean time direction. 
We have chosen the 2-direction to represent Euclidean time. When two external 
quark charges are separated along the 1-direction, the world-sheet of the
confining string extends in the space-time directions $\mu = 1,2$, while 
the string fluctuates in the transverse 3-direction. The static quark potential
$V(r)$ results from the Polyakov loop correlation function
\begin{equation}
\langle \Phi_0 \Phi_r \rangle = \frac{1}{Z} \int {\cal D}U \ \Phi_0 \Phi_r
\; \mbox{e}^{- S[U]} \sim \mbox{e}^{- \beta V(r)},
\end{equation}
in the zero-temperature limit $\beta \rightarrow \infty$. 

In order to ensure a good projection on the ground state of the string and in 
order to cover a wide range of spatial distances satisfying $r<\beta$, we have 
simulated at inverse temperatures as large as $\beta = 48$ in lattice units. 
The spatial lattice size was $L = 54$ and the bare gauge coupling was chosen as 
$4/g^2 = 9.0$ which puts the deconfinement phase transition at 
$\beta_c \approx 6$. The numerical simulations have been performed using the 
L\"uscher-Weisz technique \cite{Lue01,Lue02} with 2 levels. We 
measure the static potential using the Polyakov loop two-point function and we 
obtain $\sigma = 0.025897(15)$ for the string tension: this result is compatible with the
value reported in \cite{Cas04}.

\begin{figure}[ht]
\includegraphics[width=0.46\textwidth,clip=]{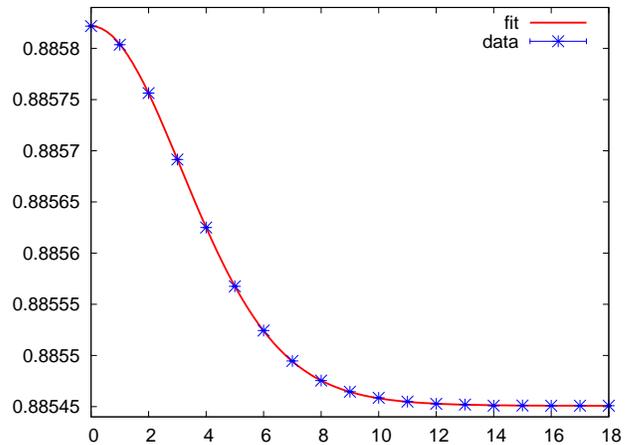}
\caption{\it The ratio 
$\langle \Phi_0 \Phi_r P_x \rangle / \langle \Phi_0 \Phi_r \rangle$
as a function of the transverse displacement $x_3$ at fixed  distance $r=19$ 
between the external static quarks. The solid line is a fit of the numerical 
data using eq.(\ref{fitbell}).}\label{bell}
\end{figure}

\begin{figure}[ht]
\includegraphics[width=0.46\textwidth]{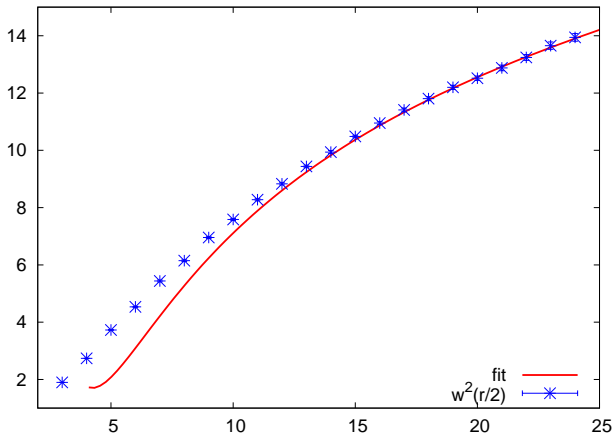}
\caption{\it The squared width of the confining string $w^2(r/2)$ at its 
midpoint as a function of the distance $r$ between the external quark charges. 
The solid curve is a fit to the next-to-leading order prediction of the low-energy 
effective field theory from eq.(\ref{widthcorr}).}\label{widthplot}
\end{figure}

The width of the fluctuating string is obtained from the connected correlation function
\begin{equation}\label{obs}
C(x_3) = \frac{\langle \Phi_0 \Phi_r P_x \rangle}{\langle \Phi_0 \Phi_r \rangle}
- \langle P_x \rangle,
\end{equation}
of a pair of Polyakov loops with a single plaquette 
\begin{equation}
P_x = \frac{1}{2}\mbox{Tr} [U_{x,1} U_{x+\hat 1,2} U_{x+\hat 2,1}^\dagger U_{x,2}^\dagger],
\end{equation}
in the $1$-$2$ plane, which measures the color-electric field along the 
1-direction of the string as a function of the transverse displacement $x_3$.
The plaquette is located at the site $x = (r/2,0,x_3)$ and thus measures the 
transverse fluctuations of the color flux tube at the maximal distance $r/2$ 
from the external quark charges. The correlation function at distance $r=19$ 
is illustrated in figure~\ref{bell}.
The data show the expected bell-shape of a Gaussian distribution, but their 
high numerical accuracy also reveals small deviations. We fit the data 
using the ansatz
\begin{equation}\label{fitbell}
\frac{\langle \Phi_0 \Phi_r P_x \rangle}{\langle \Phi_0 \Phi_r \rangle}
= A\, \exp(- x_3^2/T)\; 
\frac{1 + B\, \exp(- x_3^2/T)}{1 + D\, \exp(- x_3^2/T)} + K,
\end{equation}
where $A$, $B$, $D$, $T$, and $K$ are fit parameters. This function always provides an
excellent fit of the data. The squared width of the string is then obtained as the second
moment of the correlation function
\begin{equation}
w^2(r/2) = \frac{\int dx_3 \ x_3^2\, C(x_3)}{\int dx_3 \ C(x_3)}.
\end{equation}
In order to check the dependence of the measured string width on the ansatz of
eq.(\ref{fitbell}), we have also considered other fit functions. As long as the
ansatz provided a very good fit of the numerical data, we have not observed a
significant effect on the measured string width. Hence, systematic uncertainties
due to the fitting ansatz were negligible compared to the statistical errors. 
In figure~\ref{widthplot} we illustrate the dependence of the squared string
width $w^2(r/2)$ on the distance $r$ between the external static quarks.
At distances larger than $r \approx 18$, the Monte Carlo data are well represented by the
effective field theory prediction of eq.(\ref{widthcorr}). In particular, the predicted
prefactor $1/(2 \pi \sigma)$ of the logarithmic term $\log(r/r_0)$ is confirmed by the
numerical data. It should be noted that the string tension has already been determined by
the fit to the static quark potential. The only free parameter in the fit of the Monte
Carlo data for the width is the scale $r_0$: we obtain $r_0=2.26(2)$ such that 
$r_0 \sqrt{\sigma} = 0.364(3)$. 

It should be noted that the effective theory prediction of eq.(\ref{widthcorr}) does not
include lattice artifacts due to the violation of rotation invariance. Since the lattice
theory is invariant only under discrete rotations and not under the full Poincar\'e group,
before one reaches the continuum limit the two additional terms $\sum_{\mu = 1,2} (\p_\mu
\p_\mu h)^2$ and $\sum_{\mu = 1,2} (\p_\mu h)^4$ enter the effective theory.  Since these
terms contain four derivatives, they are of next-to-leading order. Hence, they have no
effect on the L\"uscher term or on the leading logarithmic behavior of the string width.
The corrections due to the rotation symmetry breaking terms can be computed \cite{gpw} and
included in the fit for the string width. This adds two free parameters and makes the fit
excellent down to $r \approx 10$. However, an approximate estimate of the correction to
the string width coming from the rotation invariant next-to-next-to-leading order may also
significantly improve the agreement in the range $r=10-18$. Since, at the moment, we cannot
clearly disentangle these two effects, we have not taken into account rotation symmetry
breaking corrections in our data analysis.

\begin{figure}[h]
\includegraphics[width=0.46\textwidth]{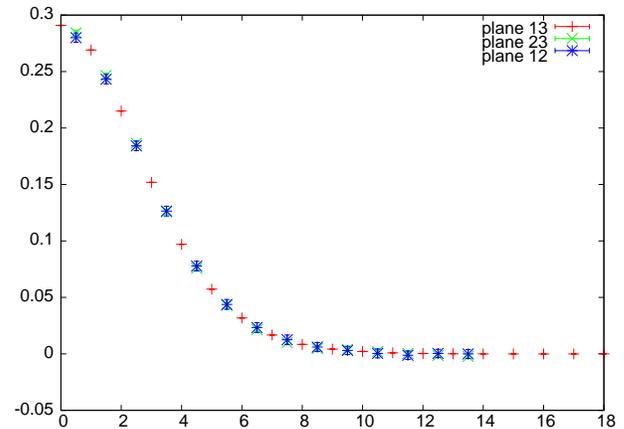}
\caption{\it Probability distribution $C(x_3)/\int dx_3 \ C(x_3)$ using the 3 
possible orientations of the plaquette in eq.(\ref{obs}). The data points for 
the 2 orientations orthogonal to the string world-sheet are put at the midpoint 
between 2 lattice sites.}\label{orientation}
\end{figure}

In the numerical results discussed so far, the plaquette orientation was parallel to the
string world-sheet. The plaquette acts as a probe that measures the fluctuations of the
confining string. The width depends on the probe through the value of the low-energy
parameter $r_0$. Choosing different probes --- and thus different ways of defining the
string width --- we expect, however, only small changes in the value of the low-energy
parameter $r_0$. In order to investigate this issue, we have considered different
orientations of the plaquette in the definition of eq.(\ref{obs}). In
figure~\ref{orientation} we show the normalized probability distribution $C(x_3)/\int dx_3
\ C(x_3)$ for the 3 different orientations of the plaquette at fixed distance $r=12$
between the external static quarks. The numerical data show that the normalized
probability distribution is not significantly affected by the different probes used to
extract the color flux tube width.  Early investigations in 4-d can be found
in~\cite{Som88}.

We gratefully acknowledge helpful discussions with M.~Caselle, P.~Hasenfratz,
F.~Niedermayer, R.~Sommer, and P.~Weisz.  This work is supported in part by funds provided
by the Schweizerischer Nationalfonds (SNF).  The computations reported in this paper have
been partially performed on the cluster {\it Lagrange} at CILEA.

\end{document}